%% file: root.tex
\documentclass[letterpaper, 10 pt, conference]{ieeeconf}  

\IEEEoverridecommandlockouts                              

\usepackage{graphics} 
\usepackage{epsfig} 
\usepackage{amsmath} 
\usepackage{amssymb}  
\usepackage{amsfonts}  
\usepackage{todonotes}
\input{ACSD-preamble}
\usepackage{xcolor}
\usepackage{subcaption}
\usepackage{siunitx}
\title{\LARGE \bf A map-based model predictive control approach for train operation}

\author{Michael Hauck, Patrick Schmidt, Alexander Kobelski, Stefan Streif
\thanks{The authors are with Technische Universit\"at Chemnitz, 09126 Chemnitz, Germany, Automatic Control and System Dynamics Lab (e-mail: \{michael.hauck, patrick.schmidt, alexander.kobelski, stefan.streif\}@etit.tu-chemnitz.de)}
\thanks{This research was funded by the German Ministry for Education and Research (BMBF) in the frame of the SRCC\_ EETCM  project, grant number 03WIR1208.}
\thanks{©\the\year \ the authors. This work has been accepted at the ECC for publication under a Creative Commons Licence CC-BY-NC-ND.
}
}
\begin{document}
\maketitle
\thispagestyle{empty}
\pagestyle{empty}

\begin{abstract}                
	Trains are a corner stone of public transport and play an important role in daily life.
A challenging task in train operation is to avoid skidding and sliding, i.e. spinning or blocking of the wheels, during fast changes of traction conditions, which can, for example, occur due to changing weather conditions, 
crossings, tunnels or forest entries. 
The latter depends on local track conditions and can be recorded in a map together with other location-dependent information like speed limits and inclination.
In this paper, a model predictive control (MPC) approach is developed.
Thanks to the knowledge of future changes of traction conditions, the approach is able to avoid short-term skidding and sliding even under fast changes of traction conditions.
In a first step, an optimal reference trajectory is determined by a multiple-shooting approach. 
In a second step, the reference trajectory is tracked by an MPC setup. 
The developed method is simulated along a track with fast-changing traction conditions for different scenarios, like changing weather conditions and unexpected delays. 
In all cases, skidding and sliding is avoided.

\end{abstract}

\section{Introduction}\label{sec:intro}

A key aspect in train operation is slip control, since high slip results in flattened wheels and track damage. 
As a result, it decreases the reliability of train operation \cite{Cai2015-novel}. 

There exist a variety of approaches for anti-skid and anti-slip control.
The majority of existing methods focus on re-adhesion control to decrease the slip after the start of skidding or sliding. 
For example, a re-adhesion control for a single-inverter multiple-induction-motors train system is presented in \cite{Matsumoto2001-novel}. 
In order to suppress undesired slipping, the motor torque is controlled. 
In \cite{Amodeo2009-wheel}, a sliding-mode observer is developed to estimate the actual traction condition, which is used as an input for a second-order sliding-mode slip control. 
In \cite{Hara2012study}, a re-adhesion slip control based on excessive torque and excessive angular momentum is presented.
However, all re-adhesion slip control approaches have the disadvantage that they react after the occurrence of undesired slipping and thus, damage of wheels and track can not be avoided. 

The emergence of undesired slipping can be tackled by active control approaches. 
In contrast to the re-adhesion or so-called "passive" slip controls, active slip controls \cite{Cai2015-novel,Hu2011-electric,Liao2014-novel} try to track the optimal adhesion value in order to avoid undesired slipping. 
The adhesion value is determined under the assumption that it depends only on slip. 
However, the influence of weather and track conditions on the adhesion is neglected or at least reduced to two possible values, which may lead to unsafe scenarios \cite{Huang2021-iterative}. 
For this reason, time-varying adhesion dynamics are introduced to tackle the changing traction conditions between wheel and track. 
A spatial iterative learning control is implemented to track the maximum adhesion. 
However, the maximum adhesion controls lead to severe impact and reduces the riding comfort~\cite{Cai2015-novel}.

Another possibility to avoid undesired slipping is to use model predictive control (MPC) approaches.
MPC is a well-known method to solve optimal control problems.
Both, input and state constraints can be incorporated easily via this method. 
In addition, updated data can be included to handle disturbances.
Predictive control algorithms in train-motion control have been previously explored for various scenarios. 
For example, in \cite{Cao2019-application}, fuzzy predictive control is considered and the authors show in simulations that the performance of train safety, comfort, and other performance indicators are improved in comparison with conventional control technology.
In \cite{Sadr2016-predictive}, a predictive field-oriented control of the induction motor is developed based on estimation of the actual traction coefficient and measurement of actual wheel slip.
A predictive speed profile tracking algorithm under linear constraints and under the assumption that the traction coefficient is constant is developed in \cite{Molavi2022-robust}.
Furthermore, in \cite{Novak2021-energy}, an energy-efficient predictive train-motion control is developed, again, under the assumption of constant traction conditions. 
There, an optimal control problem is solved backwards in time via dynamic programming, which improves the driving mode by saving up to 40\% of the consumed energy.

To the best of the authors' knowledge, existing predictive slip controls assume constant traction conditions over the prediction horizon. 
As a result, they fail to avoid skidding and sliding during fast changes of the traction conditions, which can, for example, occur during the entry in forests, tunnels or stations.  
In comparison to existing approaches in \cite{Molavi2022-robust, Novak2021-energy}, the current framework considers a variable traction coefficient, which is stored in a track-condition map together with topography data. 
Based on the knowledge of the future traction, a model predictive control approach is presented in this paper, which is able to avoid skidding and sliding even for fast changes in the traction conditions.
Recording traction information and using it for vehicle operation has already been successfully applied in other scenarios, see \cite{Kobelski2020-method, Nagase1989-study, osinenko2014adaptive, osinenko2016experimental}.

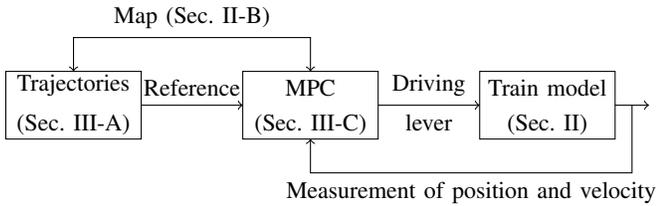
\begin{figure}	
	\centering
	\vspace{0.2em}
	\begin{tikzpicture}[scale = 0.9]
		\draw (-3.5,0)  node[above] {\small Trajectories};
		\draw (-3.5,0)  node[below] {\small (Sec. \ref{sec:methods_ref_trajs})};
		\draw (-4.5,-0.5) rectangle (-2.5,0.5);
		
		\draw (-1.75,0)  node[above] {\small Reference};
		\draw (-1.75,0)  node[below] {};
		\draw[->] (-2.5,0) -- (-1,0);
		
		\draw (0,0)  node[above] {\small MPC};
		\draw (0,0)  node[below] {\small (Sec. \ref{sec:methods_track_mpc})};
		\draw (-1,-0.5) rectangle (1,0.5);
		
		\draw (1.75,0)  node[above] {\small Driving};
		\draw (1.75,0)  node[below] {\small lever};
		\draw[->] (1,0) -- (2.5,0);
		
		\draw (3.5,0)  node[above] {\small Train model};
		\draw (3.5,0)  node[below] {\small (Sec. \ref{sec:model})};
		\draw (2.5,-0.5) rectangle (4.5,0.5);
		
		\draw (-1.75,1)  node[above] {\small Map (Sec. \ref{sec:model-map})};
		\draw[<->] (-3.5,0.5) |-  (-1.75,1) -| (0,0.5);
		
		\draw (2.375,-1)  node[below] {\small Measurement of position and velocity};
		\draw[->] (4.75,0) |-  (1.5,-1) -| (0,-0.5);
		
		\draw[->] (4.5,0) -- (5,0);
	\end{tikzpicture}
	\caption{Flow-chart of proposed map-based model predictive control approach for train operation.}
	\label{fig:setup}
	\vspace{-1.5em}
\end{figure}

The rest of the paper is organized as follows:
In Section~\ref{sec:model}, the train-motion model and acceleration constraints are introduced. 
Additionally, the track-condition-and-topography map is presented as well as the optimal control problem.
In Section~\ref{sec:methods}, a two-step solution approach is developed to solve the given problem.
Section~\ref{sec:impl-case-study} provides a case-study along with a discussion, where different situations like change of weather and unexpected delays are considered.
Finally, the paper is concluded in Section~\ref{sec:concl-outlook},  presenting an outlook on further research topics.
The whole setup is shown in Fig.~\ref{fig:setup}.

Notation: The set of component-wise non-negative vectors in $\mathbb{R}^n$ is defined as
\begin{equation*}
	\mathbb{R}_+^n := \{ z = \begin{pmatrix} z_1 & \ldots & z_n \end{pmatrix}^\top \in \R^n: z_i \geq 0, \ \forall i = 1, \ldots, n \}
\end{equation*}
for $n \in \N$.
The set of piece-wise continuous functions mapping $[0, T]$ to $\R_+$ is denoted by $\mathcal{PC}^0([0,T],\R_+)$, similar for $\mathcal{C}^k([0,T],\R_+)$ as the set of $k$ times continuously differentiable functions, where $k \geq 0$.

\section{Problem setting} \label{sec:model}

In this section, a control-oriented model of the train dynamics and restrictions of the train motion are introduced.
Afterwards, the concept of the track-condition-and-topography map, its content and how it is obtained is explained.
In the end, the optimal control problem is stated. 
	
\subsection{Train dynamics} \label{sec:model-train-dynamics}
	
The train motion can be determined by a balance of force equation \cite{Yao2018-robust}: 
\begin{equation}
	\label{eq:model_train}
	m a =  - F_\G{r}(v) +  F_\G{train}(v) u ,
\end{equation}
where $m$ is the train mass, $a$ is the acceleration, $v$ is the velocity, and $u$ is the driving-lever position of the train. 
By convention, the latter is scaled in $[-1,1]$; a positive value results in acceleration, a negative one in braking.
	
The resistance force in dependence of velocity $v$ is given as \cite{Wende2013-fahrdynamik}
\begin{equation}\label{eq:F_resistance}
	F_\G{r}(v)=  \nicefrac 1 2 \varrho_\G{air} c_\G{air}  A_\G{train} (v - v_\G{W})^2 + m g \sin(\alpha) + c_\G{R} m g
\end{equation}
and includes air resistance, inclination resistance, and rolling-resistance with $\varrho_\G{air}$ as air resistance coefficient, $c_\G{air}$ as air density, $A_\G{train}$ as effective area for train air resistance, $v_\mathrm{W}$ as wind speed in driving direction, $g$ as gravitational constant, $\alpha$ as inclination, and $c_\mathrm{R}$ as rolling resistance coefficient.
	
The driving power $F_\G{train}(v)$ is a train-specific function, depending on the maximum train power and an internal limitation for small velocities \cite{Ihme2016-schienenfahrzeugtechnik}.
Additionally, the driving power is limited by the maximum adhesion, which is addressed later by the introduction of train movement restrictions.
In \cite{Steimel2014-electric} and \cite{Beschleunigungsrechner}, the driving power is shown for different trains and it can be approximated by 
\begin{equation*}
	F_\G{train}(v) =  k_1 e^{- k_2 v} + k_3,
\end{equation*}
where the coefficients $k_i > 0$ can be determined by a least-squares approach for a specific train. 
	
Let $p$ be the position of the train and define $x = \begin{pmatrix} p & v \end{pmatrix}^\top$.
Thus, the train-motion model can be written as the following input-affine state-space model:
\begin{equation}
	\label{eq:model_train_2}
	\dot x = \underbrace{\begin{pmatrix} x_2 \\ - \frac{F_\G{r}(x_2)}{m} \end{pmatrix}}_{=: f(x; m)} + \underbrace{\begin{pmatrix} 0 \\ \frac{k_1 e^{k_2 x_2} + k_3}{m} \end{pmatrix}}_{=: g(x; m)}  u.
\end{equation}
By $f_i(x; m)$, the $i$-th component of $f(x; m)$ is denoted.
Train motion further depends on local track conditions and the topography of the track, which are summarized in a track-condition-and-topography map.
Such a map is described in the next section. 

\subsection{Track-condition-and-topography map} \label{sec:model-map}


Usually, trains travel along the same routes with location-dependent speed limits, inclination, weather-dependent traction parameters etc. 
If skidding or sliding occurs due to location-dependent reasons at a certain point, it can be assumed that this will happen again on the next trip. 
By recording such information in a track-condition-and-topography map, subsequent train rides can exploit this information to improve their performance, e.g. avoiding high slip by preemptively throttling down the engine torque.
For ease of notation, the track-condition-and-topography map is denoted as map from now on.

In this paper, we assume a standard passenger train on a commercially operated line with multiple stops.
The train must follow a timetable and adhere to speed limits. 
The track follows the topography of the terrain, so the train ride is affected by the inclination. 
Furthermore, the traction conditions of the rails are subject to environmental influences.

\begin{figure}	
	\centering
	\includegraphics[width=0.95\linewidth]{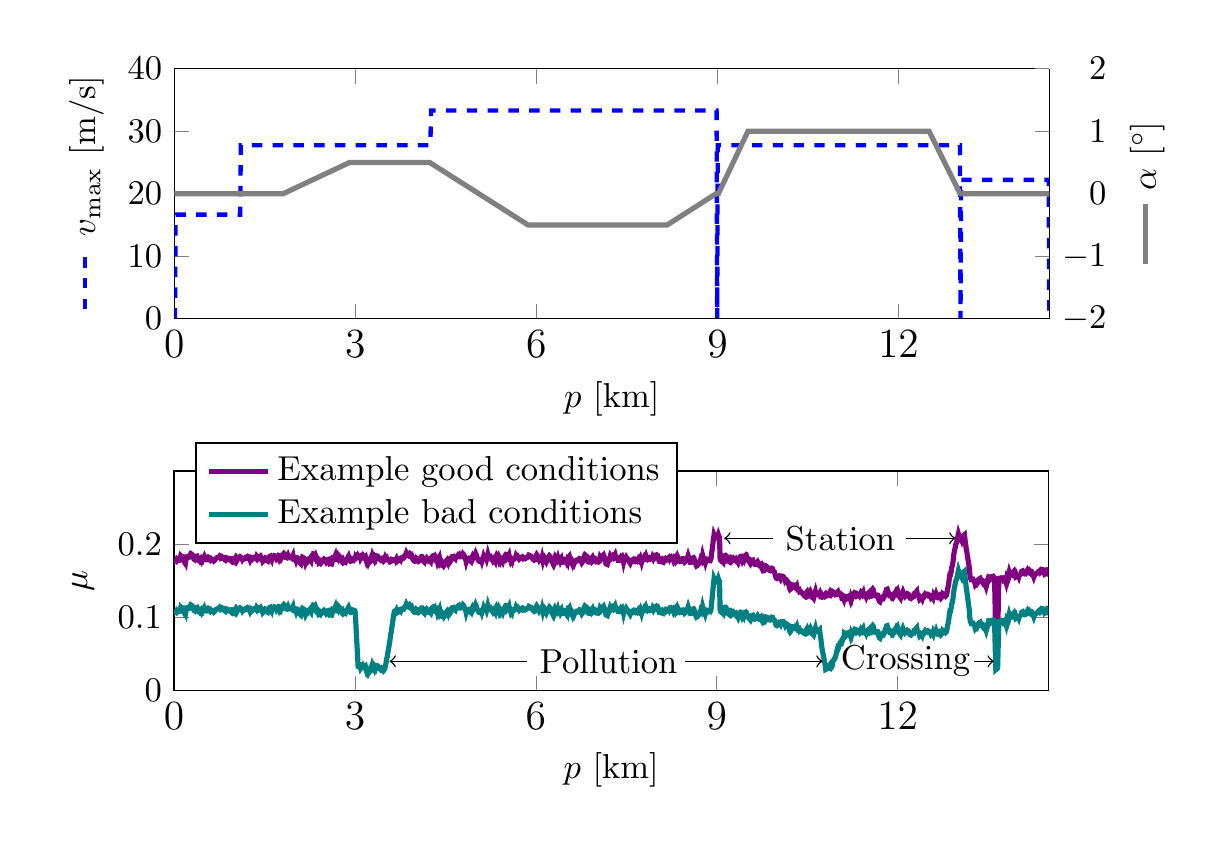}
	\vspace{-1.0em}
	\caption{Map used in case study in Section~\ref{sec:impl-case-study}. Velocity limits, inclination and time table are based on train ride from RB30 Chemnitz to Cranzahl, Germany. Traction trajectories were built based on measurements from \cite{Nagase1989-study}.}
	\label{fig:map}
	\vspace{-1.5em}
\end{figure}

For this work we assume a setup, where the map contains:
\begin{itemize}
	\item inclination at position $p$: $\alpha(p)$, 
	\item maximal velocity limit at position $p$: $v_{\G{max}}(p)$,
	\item two exemplary traction trajectories for maximal traction (also known as maximum adhesion coefficient) at position $p$: $\mu_j(p)\quad j \in \{\mathrm{good},\mathrm{bad}\}$ (based on environmental conditions), and
	\item time table including arrival time $t_{\G{station}_i}$ at station $p_{\G{station}_i}$
\end{itemize}
Two different $\mu$-trajectories will be used in the case study, to show performance on good as well as bad track conditions.
Based on this information, a map is created, see Fig.~\ref{fig:map}.
Bad conditions may refer to track conditions in freezing weather or as soon as rain begins to fall, when water mixes with dust to a muddy film on the tracks.
Note, that for future references of $\mu$, the maximum value for the given track conditions ('good' or 'bad') is denoted by $\mu_{\mathrm{max}}$.

\subsection{Constraints} \label{sec:model-constraints}

Train motion is restricted due to physical limitations (traction, engine, and brakes) as well as safety of passengers.
The acceleration is limited by the following restrictions: (i) safety restrictions for the passengers during braking and acceleration
\begin{equation}
\label{eq:const_a_max}
\abs{f_2(x; m) + g_2(x; m)u} \le a_\mathrm{max},
\end{equation} 
with $a_\mathrm{max} > 0$ and (ii) the maximum traction at the current position $p$
\begin{equation}
\label{eq:const_mu}
\begin{split}
\abs{ g_2(x; m)u } \le \mu_{\mathrm{max}}(x_1)g.
\end{split}
\end{equation}
The maximum traction $\mu_{\mathrm{max}}$ depends on the position and is defined in the next section.
Furthermore, acceleration is restricted by the maximum engine power and braking force. Both are already included in the model description with the assumption, that a driving lever position takes value between $1$ and $-1$, where $u=1$ yields maximum engine power and $u=-1$ maximum breaking force.

The total mass of the train $m$ is the mass of its own weight $m_\G{train}$ plus the load given by passengers and luggage, where the maximal additional load is defined as $m_\G{maxload}$. 
Thus, the following inequality holds for the mass of the train:
\begin{equation}
	\label{eq:const_mass}
	m_\G{min} := m_\G{train} \le m \le m_\G{train} + m_\G{maxload} =: m_\G{max}.
\end{equation}
The mass of the train varies due to a changing number of passengers and luggage.
It is constant between two stations, but unknown in most cases. 
Therefore, the constraints \eqref{eq:const_a_max} and \eqref{eq:const_mu} have to be satisfied for all $m \in [m_\G{min}, m_\G{max}]$.
Depending on whether the train is braking or accelerating, either the minimum mass or the maximum mass yields the highest value in \eqref{eq:const_a_max} and \eqref{eq:const_mu}.
Therefore, these masses are inserted into the respective inequalities.
Inequality constraints on both, states and control are incorporated via the function $h : \mathbb{R}^2 \times [-1,1] \to \mathbb{R}^7$ with 
\begin{equation} \label{eq:h}
	h(x,u) = \begin{pmatrix}
		f_2(x; m_\G{min}) + g_2(x; m_\G{min}) u - a_\mathrm{max}\\
		-(f_2(x; m_\G{max}) + g_2(x; m_\G{max}) u) - a_\mathrm{max}\\
		g_2(x; m_\G{min})u - \mu_{\mathrm{max}}(x_1)g\\
		-g_2(x; m_\G{max})u - \mu_{\mathrm{max}}(x_1)g\\
		x_2 - v_\G{max}(x_1)\\
		-1 - u\\
		u - 1
	\end{pmatrix}.
\end{equation}

\subsection{Optimal control problem} \label{sec:model-OCP}

Along with train-motion dynamics \eqref{eq:model_train}, an optimal control problem is formulated.
The solution to the optimal control problem yields the desired lever position as a control input for the model and is defined as:

\begin{equation}
\label{eq:OCP}
	\begin{split}
		u^\star = \argmin_{u \in \mathbb U} \quad & \frac{1}{2} \int_{t_0}^{t_\G{station}}u^2 \mathrm d t\\
		\mathrm{s.t.}\quad &\dot x = f(x; m) + g(x; m) u,\\
		& x(t_0) = x_0, \, x(t_\G{station}) \in \mathcal{X}_\G{station}, \\
		& h(x,u) \le 0.
	\end{split}
\end{equation}

The objective function is chosen to minimize the norm of the input within the time interval $[t_0, t_\G{station}]$, where $t_\G{station}$ is the desired arrival time at the station provided by the map.
As a solution to the optimal control problem, a continuous function $u^\star \in \mathbb U := \mathcal C^0 ([t_0, t_{\G{station}}], [-1,1])$ is obtained.
It is assumed that a feasible solution exists for all station with initial condition $x_0 \in \{ p_0 \} \times [0, v_{\G{max}}(p_0)]$, where $v_{\G{max}}(p_0)$ is the speed limit at $p_0 \in \R_+$.
As a terminal condition at $t = t_\G{station}$, the train has to stop at the station given at $p_\G{station}$.
This terminal condition is replaced by a relaxed one, which is given by $\mathcal X_\G{station}$ increasing the amount of feasible solutions.
The tolerance for the terminal position is given as $\eps_1 \geq 0$.
Therefore, the terminal set is given as  $\mathcal X_\G{station} := [p_\G{station} - \eps_1, p_\G{station} + \eps_1] \times \{ 0 \}$.

\section{Methods} \label{sec:methods}

Since the computation of $u^\star$ in \eqref{eq:OCP} is associated with high computing time, the optimal control problem has to be adjusted in order to decrease the computation burden in view of applicability.
To this end, a two-step approach is considered.
In the first step, $u^\star$ as defined in \eqref{eq:OCP} is computed to obtain a reference trajectory $u_\G{ref}^{\G{c}}:= u^\star$ for all $t \in [t_0, t_\G{station}]$. Then, $u^\star$ is applied to system \eqref{eq:model_train} to obtain the reference trajectory $x_\G{ref}^{\G{c}} \in \mathcal C^1([t_0,t_\G{station}),\R^2_+)$.
In a second step, these trajectories are tracked online via MPC. 

\subsection{Computation of reference trajectories} \label{sec:methods_ref_trajs}

The reference trajectory for the input $u$ is defined via~\eqref{eq:OCP} and it is computed with a multiple-shooting method, which is a powerful solution approach to tackle boundary value problems \cite{Bulirsch2002-introduction}.
In multiple-shooting methods, the interval $[t_0, t_\G{station}]$ is divided into $m_\ell-1$ subintervals denoted as $[t_{i}, t_{i+1}]$, $i = 1, 2, \ldots, m_\ell-1$, where $m_\ell$ is given as the number of nodes, $t_1 := t_0$, and $t_{m_\ell} := t_\G{station}$.
On each of the subintervals, the given problem is solved for initial values $x(t_i)$.
In further steps, the initial values are adjusted such that the states at the end of $[t_{i}, t_{i + 1}]$ coincide with the initial value of the next subinterval, i.e. $x(t_{i+1})$.
At the end, the boundary condition at the right-hand side of $[t_{i}, t_{i + 1}]$ and the initial condition of $[t_{i+1}, t_{i + 2}]$ are the same for $i = 1, 2, \ldots, m_\ell - 2$, which yields the desired solution $u_\G{ref}^{\G{c}} \in \mathcal{PC}^0 \left( [t_0,t_\G{station}),[-1,1] \right)$.

Once $u_\G{ref}^{\G{c}}$ is computed, it is applied to system \eqref{eq:model_train} to obtain the reference trajectory $x_\G{ref}^{\G{c}} \in \mathcal C^1([t_0,t_\G{station}),\R^2_+)$, which are tracked in the second step of the procedure, namely the model predictive control.
Since these trajectories provide a feasible solution, the prediction horizon can be reduced in order to decrease the computation time, since minimizing the distance to the trajectory ensures the satisfaction of the terminal condition.
To decrease the computational burden of the tracking MPC, the system dynamics \eqref{eq:model_train_2} are first linearized at the reference trajectories $x_\G{ref}^{\G{c}}$ and $u_\G{ref}^{\G{c}}$.
Afterwards, the linearized model is discretized for the implementation.

\subsection{Linearization and discretization} \label{sec:methods_lin_disc}

Using Taylor approximation, the linearized model with states $x_\G{lin} := x - x_\G{ref}^{\G{c}}$ and control $u_\G{lin} := u - u_\G{ref}^{\G{c}}$ reads as
\begin{equation}
	\label{eq:lin_model_train}
	\dot x_\G{lin} = \underbrace{\begin{pmatrix}
		0 & 1 \\ a_{21}(t) & a_{22}(t)
	\end{pmatrix}}_{= A_\G{lin}(t)} x_\G{lin} + \underbrace{\begin{pmatrix}
		0 \\ b_2(t)
	\end{pmatrix}}_{= b_\G{lin}(t)} u_\G{lin}.
\end{equation} 
The remaining entries are calculated as
\begin{equation*}
	a_{21}(t) = \frac{\partial f_2(x; m)}{\partial \alpha } \frac{\partial \alpha}{\partial x_1} = - g \cos(\alpha_{\G{ref}}(t)) \frac{\mathrm d \alpha}{\mathrm d x_1}
\end{equation*}
with $\alpha_{\G{ref}}(t)$ as inclination at $x^{\G{c}}_{\G{ref}_1}(t)$ and $\frac{\partial \alpha}{\partial x_1}$ given from the map, as well as
\begin{equation*}
		\begin{split}
	a_{22}(t) = &- \frac{1}{m} \varrho_\G{air} c_\G{air}  A_\G{train} (x_{\G{ref}_2}^{\G{c}}(t) - v_\G{W}) \\
	&- \frac 1 m k_1 k_2 e^{-k_2 x_{\G{ref}_2}^{\G{c}}(t)} u_{\G{ref}}^{\G{c}}(t)
\end{split}
\end{equation*}
and
\begin{equation*}
	b_2(t) = \frac{k_1 e^{-k_2 x_{\G{ref}_2}^{\G{c}}(t)} + k_3}{m}.
\end{equation*}

After linearizing the train-motion dynamics~\eqref{eq:model_train}, the obtained model \eqref{eq:lin_model_train} is discretized.
The sampled states are defined as 
\begin{equation*}
	x_\G{ref_i}^{\G{d}} = \begin{pmatrix*} x_\G{ref_i}^{\G{c}}(t_\G{step}) & x_\G{ref_i}^{\G{c}}(2t_\G{step}) & x_\G{ref_i}^{\G{c}}(3t_\G{step}) & \ldots \end{pmatrix*}^\top
\end{equation*}
for $i = 1,2$, where $u_\G{ref}^{\G{d}}$ is defined analogously. 

The resulting discrete-time model is given by
\begin{equation}
	\label{eq:discrete_lin_model_train}
	x^{\G{d}}(k+1) = A^{\G{d}}(k) x^{\G{d}}(k) + b^{\G{d}}(k) u^{\G{d}}(k),
\end{equation}
where 
\begin{equation*}
	A^{\G{d}}(k) = e^{A_\G{lin}(k t_\G{step}) \cdot t_\G{step}}
\end{equation*}
and
\begin{equation*}
	b^{\G{d}}(k) = \int_{0}^{t_\G{step}} e^{A_\G{lin}(k t_\G{step}) \cdot (t_\G{step}-\tau)} b_\G{lin}(k t_\G{step}) \ \mathrm{d} \tau.
\end{equation*}

\subsection{Tracking MPC} \label{sec:methods_track_mpc}

The linearized and discretized model is used in the optimal control problem along with the reference trajectories from Section~\ref{sec:methods_lin_disc}.
An MPC setup with moving horizon $N$ is considered to track $x_\G{ref_1}^{\G{d}}$, which results in minimizing the difference between the current position at $t = k t_{\G{step}}$ and $x_\G{ref_1}^{\G{d}}(k)$.
This difference is defined as $x^{\G{d}}_1(k)$ in \eqref{eq:discrete_lin_model_train}.
In order to increase the amount of feasible solutions, the terminal constraint in \eqref{eq:OCP} is replaced by terminal costs. 
Therefore, the cost function of the optimal control problem \eqref{eq:OCP} is enlarged by terminal costs to obtain the resulting tracking MPC as
\begin{equation}
	\label{eq:MPC}
	\begin{split}
		\min_{U} \quad & \frac 1 2 U^\top R U + \frac 1 2 X_1^\top Q X_1\\
		\mathrm{s.t.}\quad &x^{\G{d}}(k+1) = A^{\G{d}}(k) x^{\G{d}}(k) + b^{\G{d}}(k) u^{\G{d}}(k), x^{\G{d}}(0) = x_0, \\
		& h(x^{\G{d}}(k), u^{\G{d}}(k)) \le 0 \quad \forall k = j,\dots, j + N-1.
	\end{split}
\end{equation}  
In the objective function, $U, X_1 \in \R^N$ are weighted via $R, Q \in \R^{N \times N}, R, Q \succ 0$.
At time instant $j$, $X_1 = \begin{pmatrix} x_1^{\G{d}}(j) & \ldots & x_1^{\G{d}}(j+N-1) \end{pmatrix}^\top$ includes $x_\G{ref_1}^\G{d}(j), \ldots, x_\G{ref_1}^\G{d}(j+N-1)$ as reference trajectory values.
Analogously, the input vector is defined as $U = \begin{pmatrix} u^{\G{d}}(j) & \ldots & u^{\G{d}}(j+N-1) \end{pmatrix}^\top$. 

The solution of \eqref{eq:MPC} is denoted as $U^\star$.
Once the optimal control sequence is obtained, only the first element of $U^\star$ is applied to the system \eqref{eq:model_train}, which yields a new initial value for the next iteration. 
Afterwards, the horizon is shifted and, in particular, the reference values in $X_1$ are updated.
This iterative scheme is applied until the arrival station is reached.
Note that if the remaining prediction length is smaller than $N$, then $N$ is reduced to the remaining length.

\section{Implementation and case-study} \label{sec:impl-case-study}

In this section, the proposed multiple-shooting method and the tracking MPC are implemented and simulated for the RB30, traveling from Chemnitz to Cranzahl, Germany. 
Physical parameters of the train, as well as the used tuning parameters of the multiple-shooting approach and the tracking MPC are listed. 
With these parameters, three different scenarios (e.g. behavior for good traction conditions, weather change, and delays) are simulated and analyzed. 

\subsection{Parameters}

For the simulations presented in this section, the train-specific model parameters and physical constants shown in Table \ref{tab:model_parameter} are used.
They result from train data sheets, whereas $k_i$ are determined by a least-squares approach.
\begin{table}[H]
	\centering
	\caption{Train-specific model parameter.} 
	\label{tab:model_parameter}
	\begin{tabular}{|c|c|}
		\hline 
		Parameter & Value \\ 	
		\hline 
		$\varrho_{\text{air}}$ & $\SI{1.2041}{\kilogram \per \meter^3}$ \\ 
		\hline 
		$A_\G{train}$ &  $\SI{10}{\meter^2}$\\
		\hline 
		$c_\mathrm{air}$ &  $0.85$\\	
		\hline 
		$c_{\text{R}}$& $0.002$ \\
		\hline
		$k_1$ & $\SI{1.516e+05}{\kilogram \meter \per \second^2}$\\
		\hline
		$k_2$ & $\SI{0.1147}{1 \per \second}$\\
		\hline
		$k_3$ & $\SI{1.564e+04}{\kilogram \meter \per \second^2}$\\
		\hline 
		$m_\G{train}$ & $\SI{68200}{\kilogram}$ \\ 
		\hline 
		$m_\G{maxload}$ & $\SI{20500}{\kilogram}$ \\  
		\hline 
	\end{tabular}
\end{table}

The train-track-specific parameters like the speed limit $v_\G{max} (x_1)$, the inclination $\alpha(x_1) $, and the maximal traction $\mu_{\G{max}}(x_1)$ are listed in the map (see Section \ref{sec:model} and Fig. \ref{fig:map}). 
In the simulations, the wind speed in driving direction was assumed by $v_\G{w}=0$.

For the implementation of the multiple-shooting method, the number of subintervals $m_\ell$ is  determined depending on $t_\G{station}$, such that the time between two subintervals is less than three seconds, i.e. $m_\ell = \lfloor \nicefrac{t_\G{station}}{3} \rfloor$.
In the simulations it turned out to be a good trade-off between accuracy and computational effort.
However, the number of subintervals can be increased, since the computation is done offline.
For the implementation of the MPC setup, the discretization time $t_\G{step}=1$ second and the horizon length $N=20$ are used. The mass of the train $m=78200 \G{kg}$ is determined as the sum of the empty train mass and an average passenger weight value.
The weighting matrices are chosen as $R = 0.01 I_N$ and $Q = I_N$, where $I_N$ is the $N \times N$ identity matrix. 
These values are tuned manually, such that tracking of optimal states $x$ and the optimal input $u$ is fulfilled in all considered simulations.

\subsection{Simulation results}

In a first scenario, the behavior of the train is simulated and analyzed for good traction conditions. 
Afterwards, a change in the traction conditions due to fallen leaves or frozen railway resulting in a different $\mu_{\G{max}}(x_1)$ is analyzed.  
In a third simulation, the effects of an unplanned delay are investigated.

\begin{figure}	
	\includegraphics[width=0.95\linewidth]{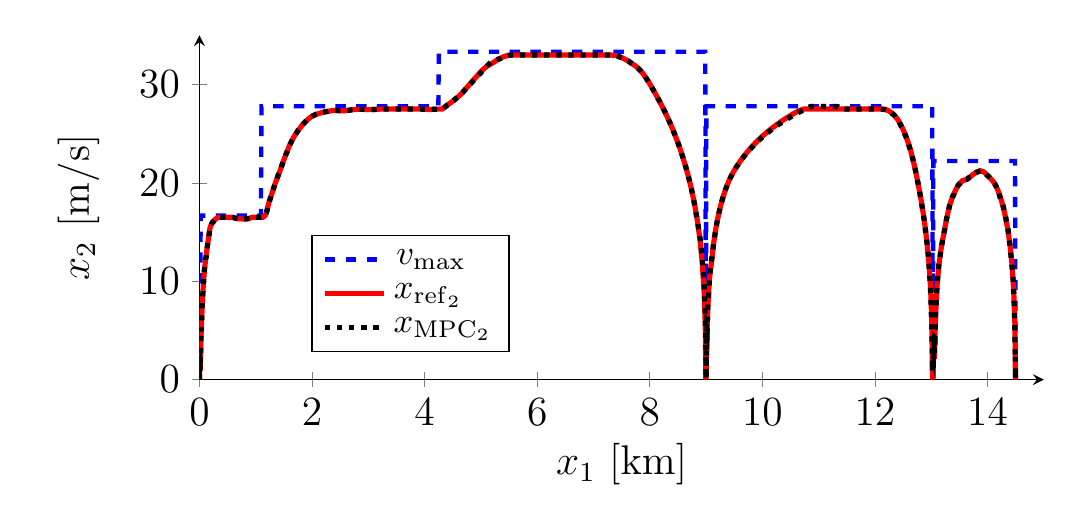}
	\vspace{-1.0em}
	\caption{Comparison of reference velocity according to multiple-shooting approach and the velocity determined by tracking MPC for the first three railway sections.}
	\label{fig:MPC_sun}
	\vspace{-1.5em}
\end{figure}

\begin{figure}
	\includegraphics[width=0.95\linewidth]{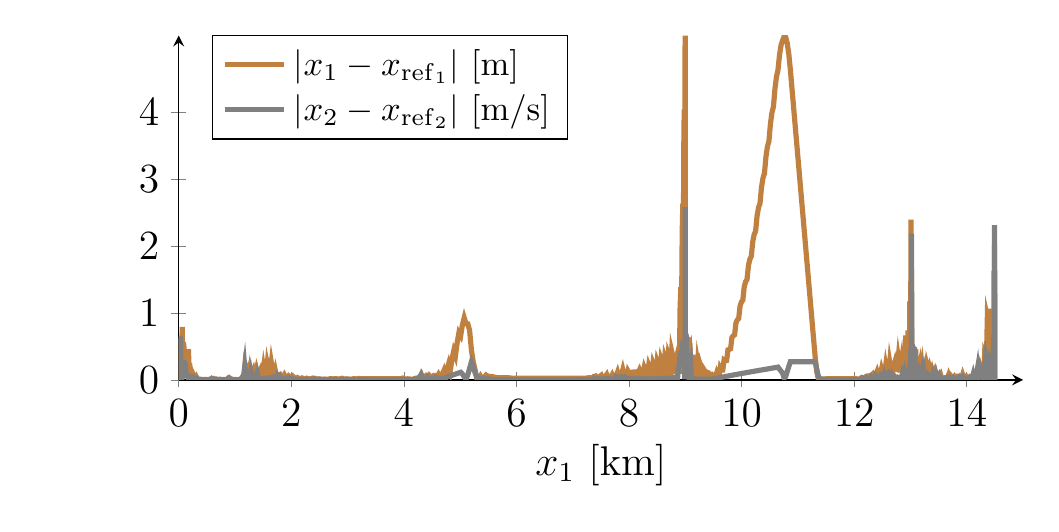}
	\vspace{-1.0em}
	\caption{Resulting tracking error between the reference trajectory $x_{\mathrm{ref}}$ and the calculated states $x$ according to tracking MPC for good traction conditions.}
	\label{fig:MPC_sun_error}
	\vspace{-1.5em}
\end{figure}

In Fig.~\ref{fig:MPC_sun}, the reference velocity according to multiple-shooting approach as well as the resulting velocity determined by tracking MPC is shown.
The difference between them is shown in Fig.~\ref{fig:MPC_sun_error} together with the resulting tracking error between the reference trajectory $x_{\mathrm{ref}_1}$ and the train position $x_1$ according to tracking MPC is shown in Fig.~\ref{fig:MPC_sun_error}.
The tracking MPC approach is able to track the reference with a maximal position error of around 5 m and a maximal velocity error of less than 3 m/s.

%

\begin{figure}	
	\includegraphics[width=0.95\linewidth]{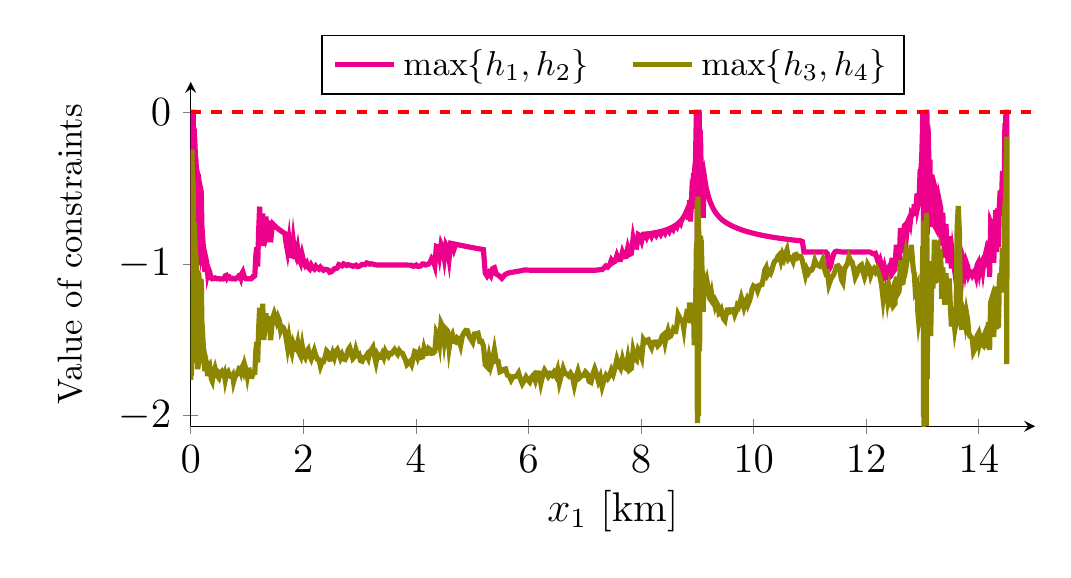}
	\vspace{-1.0em}
	\caption{Safety and traction constraints of tracking MPC with good traction conditions.}
	\label{fig:MPC_sun_constraints}
\end{figure}

Furthermore, the satisfaction of the constraints \eqref{eq:h} is considered.
The first two elements $h_1(x,u)$ and $h_2(x,u)$ describe the safety constraints, whereas $h_3(x,u)$ and $h_4(x,u)$ describe maximum traction.
In Fig.~\ref{fig:MPC_sun_constraints}, the maximum of the safety constraints as well as the traction constraints are visualized.
It can be seen that the safety constraints are more restrictive than traction constraints.
However, all constraints are satisfied since all $h_i(x,u) \leq 0$ along the whole track.

In a next step, the influence of changing traction conditions between two stations is analyzed.
There exist two possible cases, namely that the new traction conditions are better or worse than the assumed ones. 
The first case is not problematic, since the bad weather reference trajectory can be easily tracked with better traction conditions.
In the other case, a decrease of the maximal traction yields a decrease of maximal acceleration and braking. 

\begin{figure}	
	\includegraphics[width=0.95\linewidth]{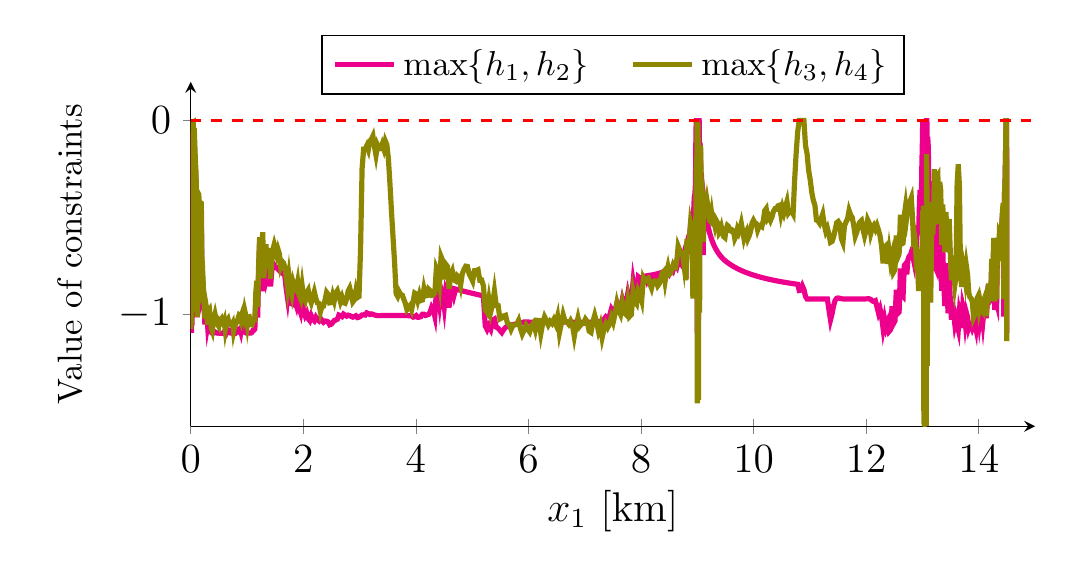}
	\vspace{-1.0em}
	\caption{Safety and traction constraints of tracking MPC with bad traction conditions.}
	\label{fig:MPC_rain_constraints}
	\vspace{-1.5em}
\end{figure}

For this scenario, the reference trajectories for good traction conditions are tracked via MPC. 
However, due to an unexpected change of the weather conditions, the traction is changing to bad conditions immediately after leaving the station (cf. traction trajectories in the map in Fig.~\ref{fig:map}). 
The tracking error yields no significant change.
More interesting is, that the traction constraints are more restrictive at the bigger part of the track, which can be seen in Fig.~\ref{fig:MPC_rain_constraints}. 
However, the constraints are satisfied along the whole track.
Therefore, the developed approach is able to avoid skidding and sliding during changes of weather conditions.

\begin{figure}	
	\includegraphics[width=0.95\linewidth]{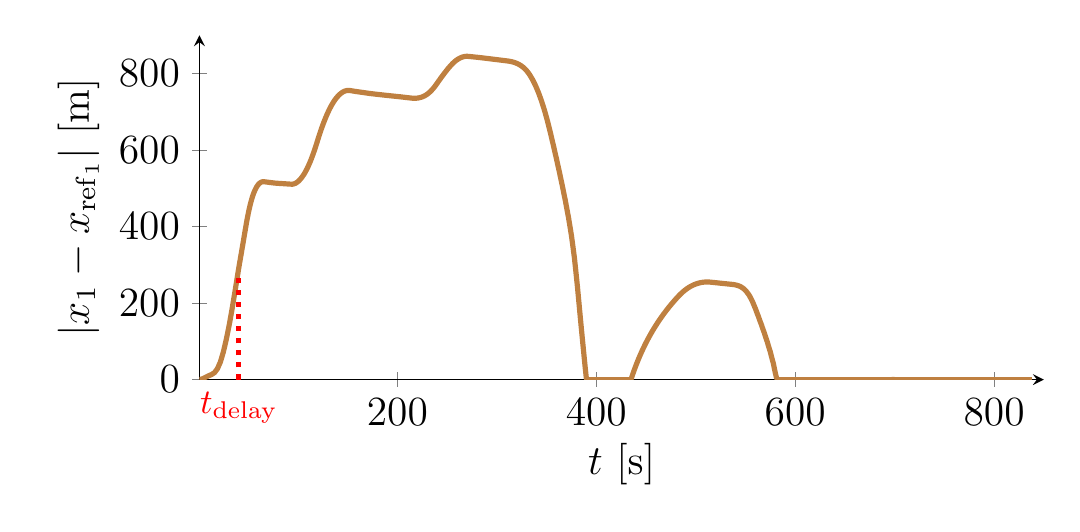}
	\vspace{-1.0em}
	\caption{Tracking error with an initial delay of $40$ seconds.}
	\label{fig:MPC_delay}
	\vspace{-1.5em}
\end{figure}

In a third scenario, the effects of an unexpected delay due to longer passenger changing time are analyzed. An initial delay of $t_\G{delay} = 40$ seconds in the first station was simulated under good traction conditions. 
With this delay, the optimal control problem \eqref{eq:OCP} is infeasible, since the train can not reach the first station in the given time.  
However, the reference trajectories are calculated offline independently of the occurred delay, thus, they are feasible for the initial time table. 
Since terminal constraints are replaced by terminal costs in the MPC approach, the OCP \eqref{eq:MPC} obtains a feasible solution.
As shown in Fig.~\ref{fig:MPC_delay}, the tracking error caused by the delay at position $x_1(t_{\G{delay}}) = 0$ is around 280 meters. 
The train arrives at the next station with a delay of 19 seconds, which is smaller than the initial one. 
Due to waiting time in the station, the tracking error is zero for a short moment until the desired departure. 
Until the following station, the tracking error decreases to zero and the MPC tracks the reference from this point on.
Again, all constraints are satisfied at each point.
Therefore, especially skidding and sliding is prevented.

\section{Conclusion and outlook} \label{sec:concl-outlook}

This paper proposes a model predictive control approach for train operation. 
The presented approach is able to avoid skidding and sliding during fast changes of the traction conditions, by using a map with traction conditions and topography data. 
In order to reduce the computational effort in online operation, the optimal control problem is solved offline to generate reference trajectories for the position, velocity and the lever position. Afterwards, a tracking MPC setup follows the reference trajectories. 

In a case-study, it turned out that the MPC tracks the reference trajectory with a small tracking error while satisfying all constraints.
Furthermore, even with changing weather conditions, MPC yields a feasible solution, where traction conditions became more restrictive in this case.
Additionally, the developed approach was able to handle unexpected delays. 

In future works, a more detailed model can be considered including, for example, motor characteristics and different braking functions.
Another future idea is an online adaption of the map based on actual measurements.
Furthermore, information that are available shortly before they have to be considered are important extensions.
Those ones include, e.g., blocked sections by other trains or environmental influences.


\bibliographystyle{bib/IEEEtran}
\bibliography{EETCM}             

                                                   

\end{document}

%% file: ACSD-preamble.tex
\usepackage{xspace}
\usepackage{soul}
\usepackage{mathtools}
\usepackage{makecell}
\usepackage{nicefrac}
\usepackage{float}
\usepackage{wrapfig}

\newcommand{\N}{\ensuremath{\mathbb{N}}}

\newcommand{\R}{\ensuremath{\mathbb{R}}}

\newcommand{\G}[1]{\mathrm{#1}}

\newcommand{\eps}{\ensuremath{\varepsilon}}

\DeclareMathOperator*{\argmin}{arg\,min}

\newcommand{\abs}[1]{\left\lvert#1\right\rvert}

\definecolor{dgreen}{rgb}{0.0, 0.5, 0.0}

\usepackage{textcomp}

\makeatletter
\newcommand{\subalign}[1]{%
	\vcenter{%
		\Let@ \restore@math@cr \default@tag
		\baselineskip\fontdimen10 \scriptfont\tw@
		\advance\baselineskip\fontdimen12 \scriptfont\tw@
		\lineskip\thr@@\fontdimen8 \scriptfont\thr@@
		\lineskiplimit\lineskip
		\ialign{\hfil$\m@th\scriptstyle##$&$\m@th\scriptstyle{}##$\crcr
			#1\crcr
		}%
	}
}
